\documentclass[doublecol]{epl2} 

\newcommand{\be}{\begin{equation}}
\newcommand{\ee}{\end{equation}}
\newcommand{\bea}{\begin{eqnarray}}
\newcommand{\eea}{\end{eqnarray}}

\newcommand{\la}{\langle}
\newcommand{\ra}{\rangle}
\newcommand{\ua}{\uparrow}
\newcommand{\da}{\downarrow}

\newcommand{\ket}[1]{\left| #1 \right\rangle}

\title{Uniqueness of density-to-potential mapping for fermionic lattice systems}
\shorttitle{Uniqueness of density-to-potential mapping for fermionic lattice systems} 

\author{J. P. Coe\inst{1} \and I. D'Amico\inst{2} \and V. V. Fran\c{c}a\inst{3}}
\shortauthor{J. P. Coe \etal}

\institute{                    
  \inst{1} Institute of Chemical Sciences, School of Engineering and Physical Sciences, Heriot-Watt University, Edinburgh, EH14 4AS, UK.\\
  \inst{2} Department of Physics, University of York, York, YO10 5DD, UK.\\
  \inst{3} Institute of Chemistry, S\~{a}o Paulo State University, Araraquara, Brazil.
}

\pacs{31.15.ec}{Hohenberg-Kohn theorem and formal mathematical properties, completeness theorems}
\pacs{71.10.Fd}{Lattice fermion models (Hubbard model, etc.)}
\pacs{71.15.Mb}{Density functional theory, local density approximation, gradient and other corrections}

\abstract{
We demonstrate that, for a fermionic lattice system, the ground-state particle density uniquely determines the external potential except for the sites corresponding to nodes of the wave function, and the limiting case where the Pauli exclusion principle completely determines the occupation of all sites.   Our fundamental finding completes, for this general class of systems, the one-to-one correspondence between ground states, their densities,
and the external potential at the base of the Hohenberg-Kohn theorem. Moreover we demonstrate that the mapping from wave function to potential is unique not just for the ground state, but also for excited states. To illustrate our findings, we  develop a practical inversion scheme to determine the external potential from a given density.  Our results hold for a general class of lattice models, which includes the Hubbard model.}

\begin{document}

\maketitle

Mapping many-body systems to a lattice while neglecting interactions beyond a given order has proven a powerful aid to devise tractable models -- such as the Hubbard \cite{OrigHubbard} and Heisenberg (see, e.g., Ref.~\cite{HeisenbergReview}) models -- and further the understanding of their properties.  Lattice systems are of recent interest for quantum chemistry calculations when the one-particle wave functions can be appropriately mapped to sites in a one-dimensional lattice and DMRG techniques are used to model the system (see, e.g., Refs.~\citenum{DMRGchem1,DMRGchem2,ChanDMRGreview2011}).  In addition, lattice systems, both bosonic and fermionic, may be simulated experimentally using an optical lattice (see, e.g., Refs.~\citenum{Zoller2005,OpticalLatticeReview,Simon2011}). The latter is of great interest to the quantum technology community for developing quantum simulators.  Fermionic lattice systems have been used to model the entanglement of nanostructures \cite{Coe11} and, using time-dependent density-functional theory, have been employed to investigate the time evolution of the out-of-equilibrium Mott insulator \cite{Karlsson11}. Time-dependent density-functional theory has also been used to create a method to simulate driven lattice gases with interactions \cite{Dierl11} and used with other classical lattice Hamiltonians to model particle transport against a bias \cite{Einax10}.  Lattice Hamiltonians have been used to describe  the process of excitation-energy transfer and the role of intramolecular vibrations in the photosynthetic process \cite{latticePhotoSynth}. Lattice systems are also used to simulate molecular systems, as for example the Pariser-Parr-Pople (PPP) model \cite{PariserParr53,Pople53}, similar to the inhomogeneous Hubbard model, can describe molecular systems with delocalised electrons in $\pi$-orbitals and has recently been used to model armchair polyacenes in Ref.~\citenum{recentPPP}.

In the last decades density-functional theory (DFT) \cite{HK} in all its flavors has been a very important tool to predict properties of materials and structures. DFT  is in principle an exact theory and allows to map many-body problems into non-interacting systems subject to a different external potential which accounts for the effects of the many-body interactions and inherits somewhat their complexity.
The important concept of DFT on a lattice (L-DFT) was introduced by Gunnarsson and Sch\"onhammer to investigate fundamental problems in DFT \cite{LDFT1,LDFT2,LDFT3}. To improve efficiency of modelling and prediction power it is now possible to use approximations in DFT to efficiently model lattice systems (see, e.g., Ref.~\citenum{Capelle13PhysRep} and references therein).

However there are still fundamental unsolved issues related to the theory itself.
The power of DFT rests on the Hohenberg-Kohn \cite{HK} and  Hohenberg-Kohn like theorems \cite{SpinDFT,CurrentDFT,PhysRevA.50.3089} where the uniqueness of the mapping between a set of basic physics functions (e.g., the ground-state wave functions, their particle densities, and external potentials) is proven.  Crucially, with respect to Hohenberg-Kohn like theorems for DFT on a lattice \cite{PhysRevB.52.2504,WU1,Godby95}, it has never so far been proven -- or disproven -- that a unique mapping between ground-state particle density (or wave function) and the external potential actually exist, leaving the basis of the theory incomplete.

In this article we prove that for  finite fermionic lattice systems, the ground-state density uniquely determines the external potential, with the exception of the sites corresponding to nodes of the wave function and the extreme case of fully filled levels (double occupancy at all sites). We relate this exception to accidental and systematic non-uniqueness according to the classification by Capelle and Vignale \cite{SDFTnonunique}. To our knowledge this fundamental finding has not been demonstrated for a lattice system before, and it completes, for this class of systems, the one-to-one correspondence between ground states, their particle densities, and the external potential
at the base of the L-DFT Hohenberg-Kohn theorem.   We note that our demonstration also shows that the mapping from wave function to potential is unique not just for the ground state, but also for excited states.

After the formal demonstration, we  develop a practical inversion scheme to determine the external potential from a given density for a lattice system and apply it  to the one-dimensional finite Hubbard model with $N$ particles to illustrate our findings. Our scheme applies to any fermionic lattice system.

The L-DFT HK theorem on a lattice \cite{PhysRevB.52.2504} proves that when dealing with non-degenerate ground-states the particle density uniquely determines the wave function on a lattice.  Starting from this we will  demonstrate that the one-to-one correspondence between the external potential and related ground state indeed holds true when considering a system on a lattice.  We will show that given a finite lattice of $L$ sites with particle number $0<N<2LM$ where $M$ is the number of bands and where, for each energy level, the allowed site states are empty, singly or doubly occupied, and the interaction and external potential are not infinite, then the external potential is uniquely determined by the wave function.

We  first consider ground-state {\it nodeless} wave functions in a single-band lattice, and extend afterwards to the case in which nodes are present, as well as to multiband lattices.
Let us assume that there are two different potentials $\sum_{j}v_j$ and $\sum_{j}v'_j$ that give the same wave function
$|\Psi\ra$. Then

\begin{equation}
\left(\hat{H}_{0}+\sum_{j}v_j\hat{n}_{j}\right)|\Psi\ra=E|\Psi\ra
\label{eq:H1}
\end{equation}

and 

\begin{equation}
\left(\hat{H}_{0}+\sum_{j}v'_j\hat{n}_{j}\right)|\Psi\ra=E'|\Psi\ra.
\label{eq:H2}
\end{equation}

Subtracting Eq.~\ref{eq:H2} from Eq.~\ref{eq:H1} gives

\begin{equation}
\sum_{j}(v_j-v'_j)\hat{n}_{j}|\Psi\ra=(E-E')|\Psi\ra
\end{equation}

which we may write as
\begin{equation}
\sum_{j}\tilde{v}_j\hat{n}_{j}|\Psi\ra=\tilde{E}|\Psi\ra.
\end{equation}

We first consider site $1$ and site $2$. The occurrence of any two types of occupation is sufficient for the following proof to hold. We first consider a wave function that contains terms, with non-zero coefficient, no matter how small, of the form $\ket{\ua\da,S,X}$ and  $\ket{S,\ua\da,X}$ where $S$ is single occupation and $X$ represents some configuration of the remaining particles into the remaining sites. This would be, for example the case for the Hubbard model ground state with the hypotheses of $N>2$, no full filling, no complete spin polarization and not-infinite potential and interaction.

We then have
\begin{equation}
\sum_{j}\tilde{v}_j\hat{n}_{j}\ket{\ua\da,S,X}=(2\tilde{v}_{1}+\tilde{v}_{2}+k)\ket{\ua\da,S,X}
\end{equation}
where $k$ includes the terms due to $X$ and 
\begin{equation}
\sum_{j}\tilde{v}_j\hat{n}_{j}\ket{S,\ua\da,X}=(\tilde{v}_{1}+2\tilde{v}_{2}+k)\ket{S,\ua\da,X}.
\end{equation}

But $|\Psi\ra$ is an eigenfunction of $\sum_{j}\tilde{v}_j\hat{n}_{j}$ so

\begin{equation}
2\tilde{v}_{1}+\tilde{v}_{2}+k=\tilde{v}_{1}+2\tilde{v}_{2}+k
\end{equation}
which gives
\begin{equation}
\tilde{v}_{1}=\tilde{v}_{2}.
\end{equation}
A similar procedure for sites $2$ and $3$, then sites $3$ and $4$ etc. leads to all the $\tilde{v}_{j}$ being equal so
\begin{equation}
\tilde{v}_{j}=c.
\end{equation}

This means that the $v_{j}$ and $v'_{j}$ may differ at most by a constant.

 A similar proof follows also by considering terms $\ket{\ua\da,0,X}$ and $\ket{0,\ua\da,X}$ or terms such as $\ket{S,0,X}$ and  $\ket{0,S,X}$. These would apply, for example to the Hubbard model with two particles or one particle respectively.  If we have half-filling with either $U\rightarrow \infty$ or complete spin polarization then we only have single occupation and the potential is not unique as would be the case for the Heisenberg model \cite{footnote1}. 

For a multiband system, for more than two particles, the proof may be extended by considering terms of the form $\ket{(B_{a})\ua\da,(B_{b})S,X}$ and  $\ket{(B_{a})S,(B_{b})\ua\da,X}$ (and similarly for different types of occupations) where we have written the occupation of the highest occupied band for sites 1 and 2, and the $B_{k}$ represent some occupation of the lower bands at a site consistent with the total number of particles. This leads to

\begin{equation}
(2+n_{a})\tilde{v}_{1}+(1+n_{b})\tilde{v}_{2}+k=(1+n_{a})\tilde{v}_{1}+(2+n_{b})\tilde{v}_{2}+k
\end{equation}
where $n_{k}$ is the density contribution due to $B_{k}$.  Hence $\tilde{v}_{1}=\tilde{v}_{2}$, so, by iterating the procedure to the other sites, we can show that $\tilde{v}_{j}=c$ for all $j$.  The proof for $N=1$ and $N=2$ follows in a similar way.

We wish now to extend the above to ground-state wave functions with nodes. For simplicity we will consider a single node at site $3$: extension to multiple nodes is straightforward. Now the procedure is carried out for site $1$ and site $2$ then site $2$ and site $4$ etc. thereby bypassing the node and again resulting in $v_{j}$ and $v'_{j}$ only differing by a constant except at the node sites where this proof does not give any constraint on the potential.

This concludes our demonstration that {\it the system density indeed determines uniquely the potential almost everywhere, with the sole exception of the wave function node sites}.
As nodes need to be isolated sites on the lattice we can then claim that {\it for all practical purposes (f.a.p.p.) the ground-state density uniquely determines the potential on the lattice}.

  We note that for the limiting case of full filling we also do not prove that there is a unique potential, but this is consistent as any potential or interaction can then be applied with no change to the site occupations. 

We classify the exceptions to uniqueness due to the Pauli exclusion principle completely determining the occupation of all sites as `accidental nonuniqueness' according to Ref.~\citenum{SDFTnonunique}. Here Capelle and Vignale use this to label non-uniqueness due to a peculiar feature of the wave function, such as complete spin polarization in the case of spin-density functional theory. We class non-uniqueness due to nodes as both accidental and systematic, in that once a node is known to occur the potential may be varied, up to a point, at the node-site to systematically construct other potentials that give the same wave function.

The above demonstration can be straightforwardly extended to show that f.a.p.p. an excited-state wave function uniquely determines the potential by using that any wave function, including excited states, uniquely determines the corresponding density.  Conversely, given the external potential (for fixed particle-particle interaction) the excited-state wave functions are uniquely determined, so a one-to-one correspondence between an excited-state wave function and the external potential is now demonstrated.  

 Hence with finite interaction and potential f.a.p.p. the ground-state density, ground-state or excited-state wave functions uniquely determine the external potential on a lattice of $L$ sites with  $0<N<2LM$ particles where $M$ is the number of bands.
To corroborate our findings we will now develop an inversion scheme with a simple recipe of taking the nodes into account. Our findings in particular mean that, when our scheme converges to a potential that reproduces a density, this is, at least for ground states and f.a.p.p., the unique potential. We will explicitly show that the inversion scheme achieves the same final potential from different initial trial potentials.

For a generic lattice Hamiltonian with an external potential $v_i$ at each site we may write
\begin{equation}
\hat{H}=\hat{H}_{0}+\sum_{j}v_j\hat{n}_{j}.
\end{equation}

Given a wave function $|\Psi\ra$ for this system,  we then have

\begin{equation}
\la\Psi| \hat{n}_{i}\hat{H} |\Psi\ra=En_{i}.
\end{equation}
where we have used $n_{i}=\la\Psi |\hat{n}_{i}|\Psi\ra$. In terms of the constituents of the Hamiltonian this is
\begin{equation}
\la \Psi | \hat{n}_{i}\hat{H}_{0} |\Psi\ra +v_{i}\la\Psi |\hat{n}_{i}^{2}|\Psi\ra +\sum_{j\neq i}v_{j} \la\Psi |\hat{n}_{i}\hat{n}_{j}|\Psi\ra =En_{i}.
\label{eq:expectnHlattice}
\end{equation}

We then rearrange Eq.~(\ref{eq:expectnHlattice}) similarly to Ref.~\citenum{CoePRA09} and use this to create an iterative scheme to find the potential which gives the density $n_{i}^{target}$ from an initial trial potential $v_{i}^{(0)}$
\begin{eqnarray}
\nonumber v_{i}^{(k+1)}=\frac{1}{\la\Psi^{(k)} |\hat{n}_{i}^{2}|\Psi^{(k)}\ra} &\big(&n_{i}^{target}E^{(k)}-\la \Psi^{(k)} | \hat{n}_{i}\hat{H}_{0} |\Psi^{(k)}\ra\\
 &-&\sum_{j\neq i}v_{j}^{(k)} \la\Psi^{(k)} |\hat{n}_{i}\hat{n}_{j}|\Psi^{(k)}\ra\big). \label{eq:schemecomp}
\end{eqnarray}
Using the identity
\begin{eqnarray}
\nonumber \la \Psi^{(k)} | \hat{n}_{i}\hat{H}_{0} |\Psi^{(k)}\ra +\sum_{j\neq i}v_{j}^{(k)} \la\Psi^{(k)} |\hat{n}_{i}\hat{n}_{j}|\Psi^{(k)}\ra=\\-v^{(k)}_{i}\la\Psi^{(k)} |\hat{n}_{i}^{2}|\Psi^{(k)}\ra+E^{(k)}n^{(k)}_{i},
\end{eqnarray}
we may rewrite Eq.~(\ref{eq:schemecomp}) in the simpler form
\begin{equation}
\nonumber v_{i}^{(k+1)}=\frac{\left(n_{i}^{target}-n_{i}^{(k)}\right)E^{(k)}}{\la\Psi^{(k)} |\hat{n}_{i}^{2}|\Psi^{(k)}\ra} +v_{i}^{(k)}.
\end{equation}
Following arguments similar to  Ref.~\citenum{CoePRA09}, we note that this is expected to converge for $E<0$, and so define the general expression
\begin{equation}
\nonumber v_{i}^{(k+1)}=\frac{\left(n_{i}^{(k)}-n_{i}^{target}\right)|E^{(k)}|}{\la\Psi^{(k)} |\hat{n}_{i}^{2}|\Psi^{(k)}\ra} +v_{i}^{(k)}.
\label{eq:schemeHub}
\end{equation}
At convergency $v_{i}^{(k+1)}=v_{i}^{(k)}$ is the external potential that reproduces the target density via the many-body Schr\"{o}dinger equation. We note that this approach can be applied to any of the system eigenstates, including the ground-state.
 Our result Eq.~(\ref{eq:schemeHub}) is similar to that derived for continuous spatial co-ordinates in Ref.~\citenum{CoePRA09}  {\it except} for the change in the denominator. We will numerically implement this scheme with $80\%$ mixing of the previous potential to reduce the chance of instabilities.

 We now apply our scheme to a finite lattice system with fixed and finite  particle-particle interaction to find the unique potential, and hence the wave function, corresponding to a density.  To explore the behavior of the scheme at the nodes, and for practical examples of its implementation, we consider the one-dimensional Hubbard model (HM)
\begin{eqnarray}
\nonumber H_{HM}=-t\sum_{i,\sigma}\left (c_{i,\sigma}^{\dagger}c_{i+1,\sigma}+c_{i+1,\sigma}^{\dagger}c_{i,\sigma} \right)\\
+\tilde{U}\sum_{i}\hat{n}_{i,\uparrow}\hat{n}_{i,\downarrow}+\sum_{i}v_i\hat{n}_{i}.
\label{eqn:HubbardHamiltonian}
\end{eqnarray}
We use open boundary conditions, $L$ sites, interaction strength of $U=\tilde{U}/t$ and $N$ particles. 
 We first demonstrate that we can achieve the same final potential, up to an additive constant, independent of the starting potential.  For the exact system we use a harmonic potential 
\begin{equation}
v_{i}=\frac{1}{2}k\left(i-\frac{(L+1)}{2}\right)^{2}
\end{equation} 
with $k=1$ to generate the target density. We then apply the iterative scheme (\ref{eq:schemeHub}) starting from a harmonic potential with $k=2$ and then starting from $k=0.1$. Here we require a mean absolute difference of $10^{-8}$ between the trial and target site densities for convergence.

\begin{figure}[ht]\centering
\includegraphics[width=.45\textwidth]{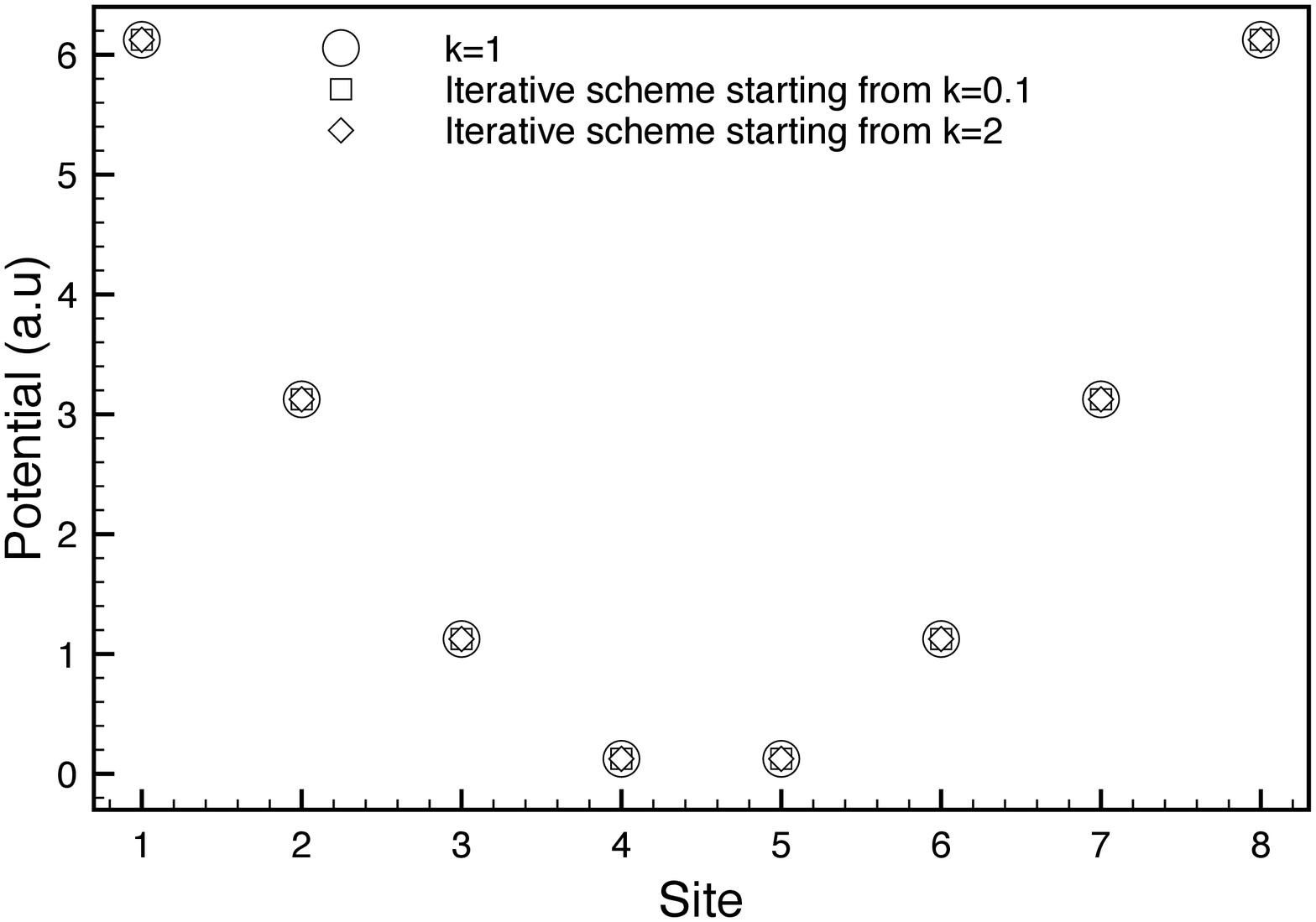}
\includegraphics[width=.45\textwidth]{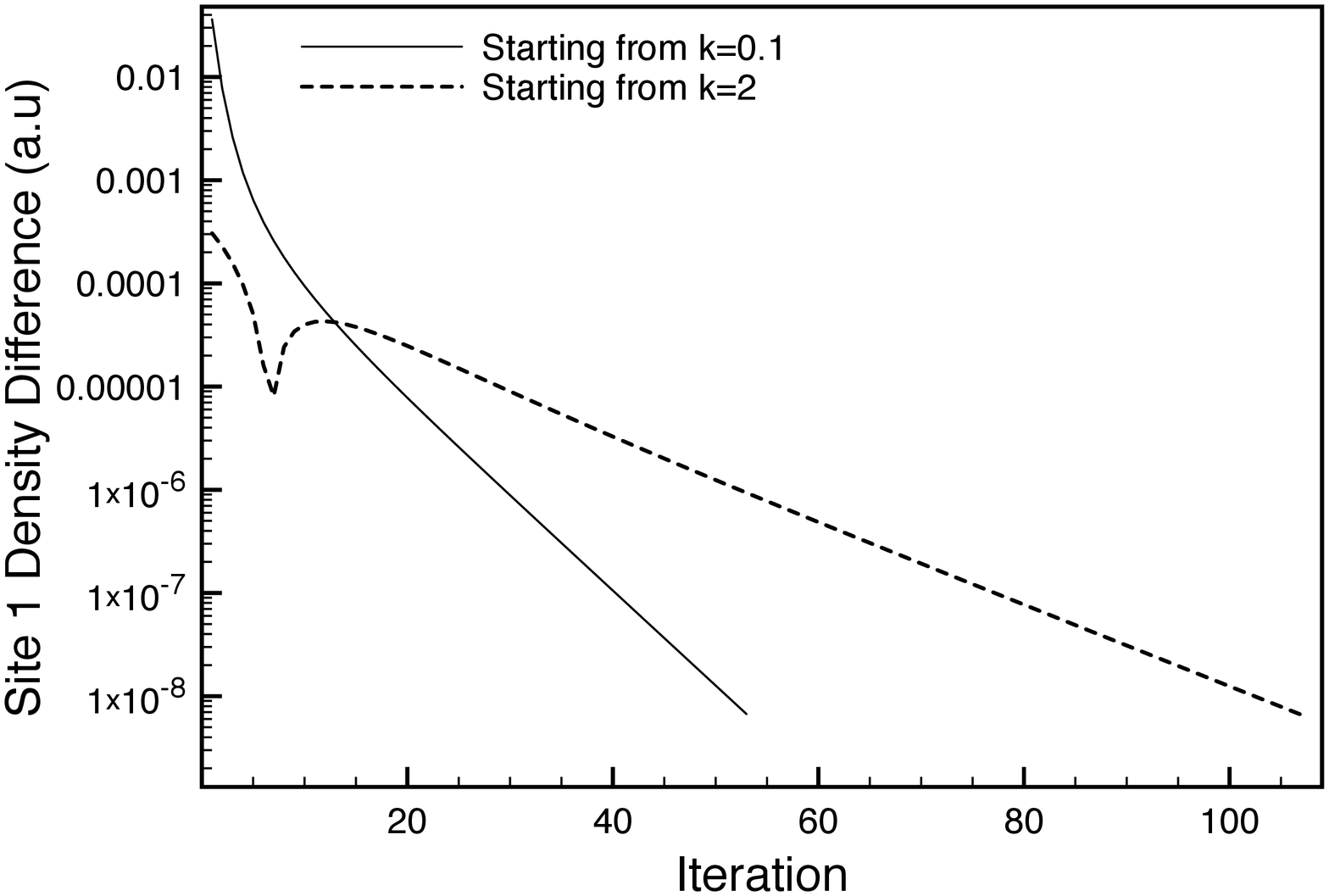}
\caption{Top Panel: Potentials, shifted by constants, found with initial trial harmonic potentials of $k=0.1$ or $k=2$ compared with the exact harmonic potential $k=1$. Here $L=8$, $N=2$ and $U=2$. Lower Panel: Absolute difference between the exact density and the density from the scheme on each iteration for site $1$.}\label{fig:varyTrial}
\end{figure}
We see in the top panel of Fig.~\ref{fig:varyTrial} that the same final potential is found, up to an additive constant, regardless of the starting potential. Fig.~\ref{fig:varyTrial} (lower panel) shows that the rate of convergence, however, may be affected by the trial potential.

 To investigate the use of the scheme when nodes are present we consider excited states to produce nodes with a finite potential.
 Our argument does not show what values the potential takes at a site with a node, so in practice,  we set the potential at these sites to zero at all iterations of Eq.~(\ref{eq:schemeHub}). We then attempt to use the scheme to find a finite potential, if any, that gives the required density.  In the calculations we prevent the possibility of the potential becoming too large, and possibly tending towards an infinite potential solution, by creating a random potential at all sites with maximum value of $10$ if any of the sites has a potential greater than $1000$. 

\begin{figure}[h]\centering
\includegraphics[width=.45\textwidth]{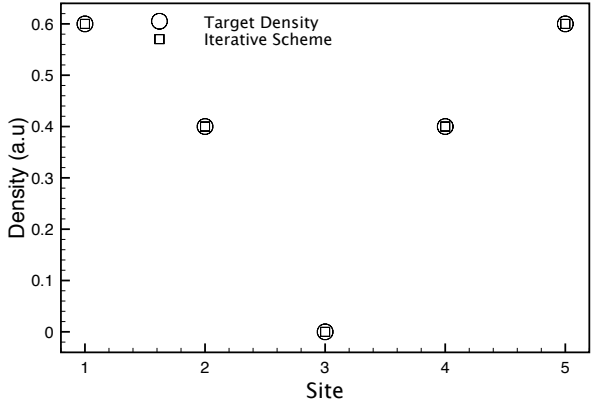}
\caption{Density from the third excited state of a system with $L=5$, $N=2$ and $U=0$ found  using the scheme (\ref{eq:schemeHub}) and compared with a target density with a node.}\label{fig:excitednode}
\end{figure}

 We first consider the case of a target density with a node at a single site (see Fig.~\ref{fig:excitednode}), where the potential is not known beforehand, and consider $U=0$.  Here our scheme indeed converges towards a potential such that  the third excited state of the system reproduces the target density with a node at site $3$, as demonstrated in Fig.~\ref{fig:excitednode}. We note that in this case the final potential at site $3$ is not unique but nor is it arbitrary: by direct solution of the Schr\"{o}dinger equation, we find that the potential can be lowered at site $3$ to around $-0.2$ before the density changes.

\begin{figure}[h]\centering
\includegraphics[width=.45\textwidth]{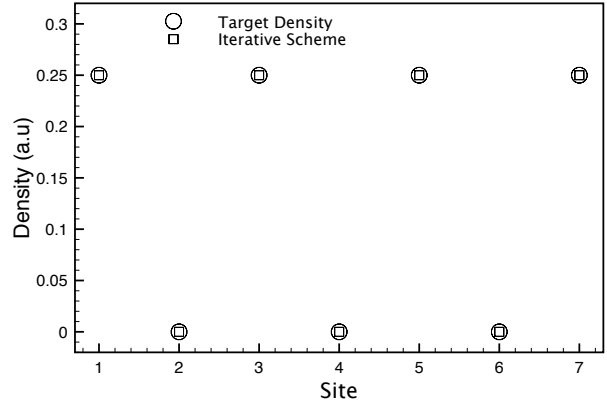}
\caption{ Target density from the third excited state of a system with $L=7$, $N=1$ and $k=0$ compared with the results from the scheme (\ref{eq:schemeHub}) when starting from a trial  harmonic potential with $k=1$.}\label{fig:3nodes}
\end{figure}

The approach can also be applied to a density with multiple nodes. To find a density of this form that could be created from a known finite potential we considered the exact solution for one particle with a constant potential.  This system with seven sites has three nodes in its  third excited state. Fig.~\ref{fig:3nodes} demonstrates that we can indeed recreate this density by finding a constant potential, albeit a different constant, when using the scheme with the $k=1$ harmonic potential as the initial guess. 

We proved that for a single or multiband fermionic lattice system without full filling nor fully polarised at half-filling the external potential is {\it uniquely} determined by the ground-state wave function except at sites with zero density nodes and hence by the ground-state density due to the unique mapping between ground-state densities and wave functions \cite{PhysRevB.52.2504}. We completed in this way the demonstration of the Hohenberg-Kohn-like theorem for Lattice-DFT. We also demonstrated a one-to-one correspondence between external potential and excited states. As a practical application, we have adapted the inversion scheme of Ref.~\citenum{CoePRA09} for a generic lattice system and applied it to  the Hubbard model. We have shown that with this scheme the external potential corresponding to a lattice density, and given the other constituents of the Hamiltonian, may be found.    Our results will prove useful for further fundamental work in density-functional theory and also in the testing, construction and improvement of approximations in lattice density-functional theory with potential applications in all fields in which lattice Hamiltonians are used. Future work will consider extending these results to finite temperature density-functional theory\cite{Mermin65,Eschrig10} for lattice systems \cite{Xianlong12}.

\acknowledgments
IDA gratefully acknowledge support from
a University of York-FAPESP combined grant.  VVF acknowledges support from FAPESP and CNPq.  We are thankful to Klaus Capelle for useful discussions.


\begin{thebibliography}{10}
\bibitem{footnote1}
In this case a spin-dependent potential should be considered.
\end{thebibliography}

\end{document}